# Isometry on Interval-valued Fuzzy Graphs

*Hossein Rashmanlou*[1] and *Madhumangal Pal*[2]

[1]Department of Mathematics, University of Mazandaran, Babolsar, Iran
Email: hrashmanlou@yahoo.com
[2]Department of Applied Mathematics with Oceanology and Computer Programming
Vidyasagar Univesity, Midnapore-721102, West Bengal, India
Email: mmpalvu@gmail.com

***Abstract.*** Theoretical concepts of graphs are highly utilized by computer scientists. Especially in research areas of computer science such as data mining, image segmentation, clustering image capturing and networking. The interval-valued fuzzy graphs are more flexible and compatible than fuzzy graphs due to the fact that they allowed the degree of membership of a vertex to an edge to be represented by interval valued in [0,1] rather than the crisp real values between 0 and 1. In this paper, we defined isometry on interval–valued fuzzy graphs and show that isometry on interval–valued fuzzy graphs is an equivalence relation.

## 1. Introduction

At present, graph theoretical concepts are highly utilized by computer science applications. Especially in research areas of computer science including data mining, image segmentation, clustering, image capturing networking, for example, a data structure can be designed in the form of tree which in turn utilized vertices and edges. Similarly modeling of network topologies can be done using graph concepts.

In 1975, Zadeh [22] introduced the notion of interval –valued fuzzy sets as an extension of fuzzy sets [21] in which the values of the membership degrees are intervals of numbers instead of the numbers. Interval–valued fuzzy sets provide a more adequate description of uncertainly than traditional fuzzy sets. It is therefore important to use interval–valued fuzzy sets in application, such as fuzzy control. Since interval– valued fuzzy sets are widely studied and used, we describe briefly the work of Gorzalczany on approximate reasoning [6,7], Roy and Biswas on medical diagnosis [13] and Mendel on intelligent control [10].

In 1975, Rosenfeld [14] discussed the concept of fuzzy graphs whose basic idea was introduced by Kauffman [9] in 1973. The fuzzy relation between fuzzy sets were also considered by Rosenfeld and he developed the structure of fuzzy graphs, obtain analogs of several graph theoretical concepts. Bhattacharya [4] gave some remarks on fuzzy graphs. Mordeson and Peng [12] introduced some operations on fuzzy graphs. The complement of a fuzzy graph was defined by Mordeson [11]. Bhutani and Rosenfeld introduced the concept of M-strong fuzzy graphs in [5] and studied some of their properties. Shannon and Atanassov [15] introduced the concept of intuitionistic fuzzy relations and intuitionistic fuzzy graphs. Hongmei and Lianhua gave the definition of interval– valued graph in [8]. Recently Akram introduced the concepts of bipolar fuzzy graphs and interval–valued fuzzy graphs in [1,2,3]. Samanta, Pal and Rashmanlou [24-33] produced several results on fuzzy graph theory. In this paper, we present the concepts of neighbourly irregular interval–valued fuzzy graphs, neighbourly totally irregular interval–valued fuzzy graphs, highly irregular interval – valued fuzzy graphs, and highly totally irregular interval–valued fuzzy graphs are introduced and investigated. A necessary and sufficient condition under which neighbourly irregular and highly irregular interval–valued fuzzy graphs are equivalent is discussed.

## 2. Preliminaries

In this section, we review some elementary concepts that are necessary for this paper.

By a graph, we mean a pair $G^* = (V, E)$, where $V$ is the set and $E$ is a relation on $V$. The elements of $V$ are vertices of $G^*$ and the elements of $E$ are edges of $G^*$. We write $xy \in E$ to mean $(x, y) \in E$, and if $e = xy \in E$, we say $x$ and $y$ are adjacent. Formally, given a graph $G^* = (V, E)$, two vertices $x, y \in V$ are said to be neighbours, or adjacent nodes, if $xy \in E$. The number of edges, the cardinality of $E$, is called the size of graph and denoted by $|E|$. The number of vertices, the cardinality of $V$, is called the order of graph and denoted by $|V|$.

A path in a graph $G^*$ is an alternating sequence of vertices and edges $v_{\circ}, e_1, v_1, e_2, ..., v_{n-1}, e_n, v_n$. The path graph with $n$ vertices is denoted by $P_n$. A path is sometime denoted by $P_n : v_{\circ} v_1 ... v_n (n > \circ)$. The length of a path $P_n$ in $G^*$ is $n$. A path $P_n : v_{\circ} v_1 .... v_n$ in $G^*$ is called a cycle is $v_{\circ} = v_n$ and $n \geq 3$. Note that path graph, $P_n$ has $n-1$ edges and can be obtain from cycle graph, $C_n$, by removing any edge. The neighbourhood of a vertex $v$ in a graph $G^*$ is the induced subgraph of $G^*$ consisting of all vertices. The neighbourhood is often denoted $N(v)$. The degree deg($v$) of vertex $v$ is the number of edges incident on $V$ or equivalently, $\deg(v) = |N(v)|$. The set of neighbours called a (open) neighbourhood $N(v)$ for a vertex $v$ in a graph $G^*$, consists of all vertices adjacent to $v$ but not including $v$, that is $N(v) = \{u \in V \mid uv \in E\}$. When $v$ is also included, it is called a closed neighbourhood $N[v]$, that is $N[v] = N(v) \cup \{v\}$. A regular graph is a graph where each vertex has the some number of neighbours, i.e. all the vertices have the same closed neighbourhood degree.

An interval number D is an interval $[a^-, a^+]$ with $\circ \leq a^- \leq a^+ \leq 1$. The interval $[a, a]$ is identified with the number $a \in [\circ, 1]$. $D[\circ, 1]$ denotes the set of all interval numbers. For interval numbers $D_1 = [a_1^-, b_1^+]$ and $D_2 = [a_2^-, b_2^+]$, we have

- $\mathrm{r}\min(D_1, D_2) = \mathrm{r}\min([a_1^-, b_1^+], [a_2^-, b_2^+]) = [\min\{a_1^-, a_2^-\}, \min\{b_1^+, b_2^+\}]$
- $\mathrm{r}\max(D_1, D_2) = \mathrm{r}\max([a_1^-, b_1^+], [a_2^-, b_2^+]) = [\max\{a_1^-, a_2^-\}, \max\{b_1^+, b_2^+\}]$,
- $D_1 + D_2 = [a_1^- + a_2^- - a_1^-.a_2^-, b_1^+ + b_2^+ - b_1^+.b_2^+]$,
- $D_1 \leq D_2 \Leftrightarrow a_1^- \leq a_2^-$ and $b_1^+ \leq b_2^+$,
- $D_1 = D_2 \Leftrightarrow a_1^- = a_2^-$ and $b_1^+ = b_2^+$,
- $D_1 < D_2 \Leftrightarrow D_1 \leq D_2$ and $D_1 \neq D_2$
- $KD = Ka_1^-, b_1^+ = [Ka_1^-, Kb_1^+]$, where $\circ \leq K \leq 1$.

Then, $(D[\circ, 1], \leq, \vee, \wedge)$ is a complete lattice with $[\circ, \circ]$ as the least element and $[1, 1]$ as the greatest.

The interval – valued fuzzy set $A$ in $V$ is defined by
$$A = \{(x, [\mu_{A^-(x)}, \mu_{A^+(x)}]) | x \in V\},$$
where $\mu_{A^-(x)}$ and $\mu_{A^+(x)}$ are fuzzy subsets of $V$ such that $\mu_{A^-(x)} \leq \mu_{A^+(x)}$ for all $x \in V$.
If $G^* = (V, E)$ is a graph, then by an interval–valued fuzzy relation $B$ on a set $E$ we mean an interval-valued fuzzy set such that
$$\mu_{B^-(xy)} \leq \min(\mu_{A^-(x)}, \mu_{A^-(y)}), \quad \mu_{B^+(xy)} \leq \min(\mu_{A^+(x)}, \mu_{A^+(y)}), \text{ for all } xy \in E.$$

**Definition 2.1.** By an interval – valued fuzzy graph of a graph $G^* = (V, E)$ we mean a pair $G = (A, B)$, where $A = [\mu_{A^-}, \mu_{A^+}]$ is an interval – valued fuzzy set on $V$ and $B = [\mu_{B^-}, \mu_{B^+}]$ is an interval – valued fuzzy relation on $E$ such that
$$\mu_{B^-(xy)} \leq \min(\mu_{A^-(x)}, \mu_{A^-(y)}), \quad \mu_{B^+(xy)} \leq \min(\mu_{A^+(x)}, \mu_{A^+(y)}).$$
Throughout in this paper, $G^*$ is a crisp graph, and $G$ is an interval – valued fuzzy graph.

**Definition 2.2.** The number of vertices, the cardinality of $V$, is called the order of an interval–valued fuzzy graph $G = (A, B)$ and denoted by $|V|$ (or $O(G)$), and defined by
$$O(G) = |V| = \sum_{x \in V} \frac{1 + \mu_{A^-(x)} + \mu_{A^+(x)}}{2}.$$
The number of edges, the cardinality of $E$, is called the size of an interval – valued fuzzy graph $G = (A, B)$ and denoted by $|E|$ (or $S(G)$), and defined by
$$S(G) = |E| = \sum_{xy \in E} \frac{1 + \mu_{B^-(xy)} + \mu_{B^+(xy)}}{2}.$$

**Definition 2.3.** Let $G$ be an interval – valued fuzzy graph. The neighbourhood of a vertex $x$ in $G$ is defined by $N(x) = (N_\mu(x), N_\nu(x))$, where
$N_\mu(x) = \{y \in V : \mu_{B^-(xy)} \leq \min(\mu_{A^-(x)}, \mu_{A^-(y)})\}$ and
$N_\nu(x) = \{y \in V : \mu_{B^+(xy)} \leq \min(\mu_{A^+(x)}, \mu_{A^+(y)})\}$.

**Definition 2.4.** Let $G$ be an interval – valued fuzzy graph. The neighbourhood degrees of vertex $x$ is $G$ is defined by $\deg(x) = (\deg_\mu(x), \deg_\nu(x))$, where
$\deg_\mu(x) = \sum_{y \in N(x)} \mu_{A^-(y)}$ and $\deg_\nu(x) = \sum_{y \in N(x)} \mu_{A^+(y)}$. Notice that
$\mu_{B^-(xy)} > \circ, \mu_{B^+(xy)} > \circ$ for all $xy \in E$, and $\mu_{B^-(xy)} = \mu_{B^+(xy)} = \circ$ for all $xy \notin E$

**Definition 2.5.** Let $G$ be an interval – valued fuzzy graph on $G^*$. If there is a vertex where is adjacent to vertices with distinct neighbourhood degrees, then $G$ is called an irregular interval–valued fuzzy graph. That is, $\deg(x) \neq n$ foa all $x \in V$.

**Example 2.6.** Consider a graph $G^* = (V, E)$ such that $V = \{u_1, u_2, u_3\}$, $E = \{u_1u_2, u_2u_3, u_3u_1\}$. Let A be an interval–valued fuzzy subset of $V$ and let $B$ be an interval– valued fuzzy subset of $E \subseteq V \times V$ defined by

|  | $u_1$ | $u_2$ | $u_3$ |
|---|---|---|---|
| $\mu_{A^-}$ | 0.3 | 0.3 | 0.4 |
| $\mu_{A^+}$ | 0.7 | 0.8 | 0.5 |

|  | $u_1 u_2$ | $u_2 u_3$ | $u_3 u_1$ |
|---|---|---|---|
| $\mu_{B^-}$ | 0.2 | 0.3 | 0.2 |
| $\mu_{B^+}$ | 0.3 | 0.4 | 0.3 |

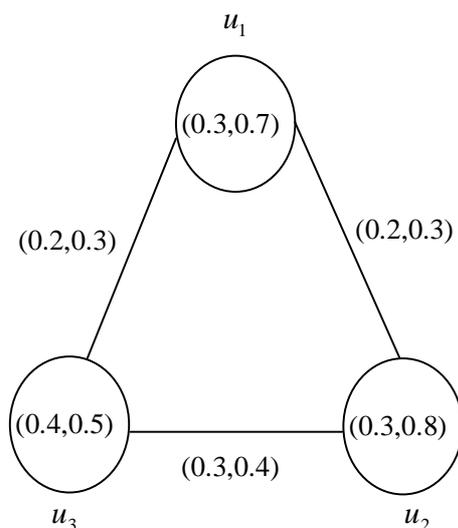

**Figure 1:** Irregular interval-valued fuzzy graph G

By routine computations, we have $\deg(u_1) = (0.7,1.3)$, $\deg(u_2) = (0.7,1.2)$ and $\deg(u_3) = (0.6,1.5)$. It is clear that $G$ is an irregular interval–valued fuzzy graph.

**Definition 2.7.** Let $G$ be an interval–valued fuzzy graph. The closed neighbour-hood degree of a vertex $x$ is defined by $\deg[x] = (\deg_\mu[x], \deg_\nu[x])$, where $\deg_\mu[x] = \deg_\mu(x) + \mu_{A^-}(x)$, $\deg_\nu[x] = \deg_\nu(x) + \mu_{A^+}(x)$.
If there is a vertex which is adjacent to vertices with distinct closed neighbourhood degrees, then $G$ is called a totally irregular interval–valued fuzzy graph.

**Example 2.8.** Consider an interval–valued fuzzy graph $G$ such that
$V = \{u_1, u_2, u_3, u_4, u_5\}$, $E = \{u_1u_2, u_2u_3, u_2u_4, u_3u_1, u_3u_4, u_4u_1, u_4u_5\}$.

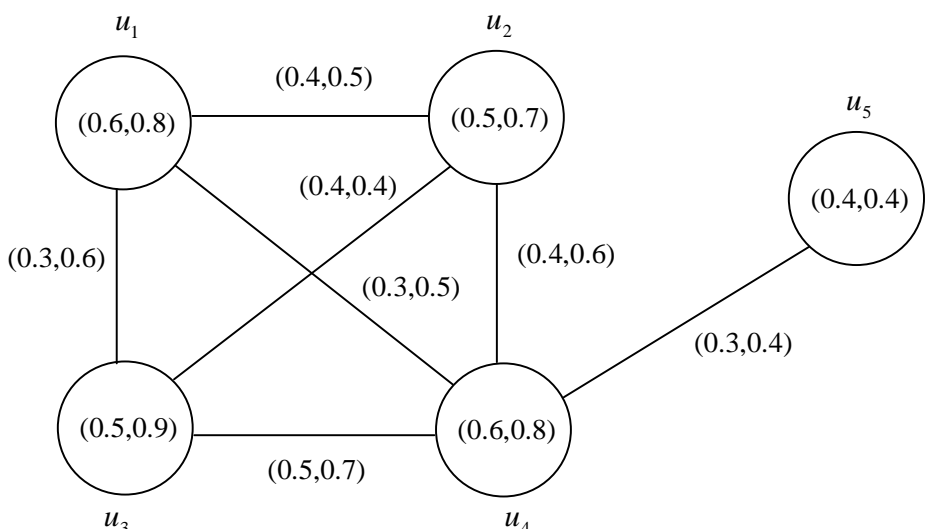

**Figure 2:** Totally irregular interval-valued fuzzy graph G

By routine computations, we have $\deg[u_1] = (2.2, 3.2), \deg[u_2] = (2.2, 3.2)$, $\deg[u_3] = (2.2, 3.2), \deg[u_4] = (2.6, 3.6), \deg[u_5] = (1, 1.2)$. It is clear from calculations that $G$ is a totally irregular interval–valued fuzzy graph.

### 3. Isometric interval-valued fuzzy graphs

**Definition 3.1.** Let $G_1 = (A_1, B_1)$ and $G_2 = (A_2, B_2)$ be interval-valued fuzzy graphs. Then $G_2$ is said to be isometric from $G_1$ if for each $v \in V_1$ there is a bijective $g_v : V_1 \to V_2$ such that $\delta_1^-(u,v) = \delta_2^-(g_v(u), g_v(v))$, $\delta_1^+(u,v) = \delta_2^+(g_v(u), g_v(v))$ for every $u \in V_1$, in which $\delta_i^-$ ($i = 1,2$) and $\delta_i^+$ (i=1,2) are $\mu_{B_i}^-$ distance and $\mu_{B_i}^+$ distance, respectively. In addition we define

Isometry on Interval-valued Fuzzy Graphs

$$\delta^-(u,v) = \wedge \sum_{i=1}^{n} \frac{1}{\mu_B^-(u_{i-1}u_i)} \text{ and } \delta^+(u,v) = \wedge \sum_{i=1}^{n} \frac{1}{\mu_B^+(u_{i-1}u_i)}, \text{ where}$$

$u = u_0, u_1, \ldots, u_i, \ldots, u_n = v$ is a path from $u$ to $v$.

**Proposition 3.2.** Let $G_1 = (A_1, B_1)$ and $G_2 = (A_2, B_2)$ be two interval-valued fuzzy graphs. Then $G_1$ is isomorphic to $G_2$ implies $G_1$ is isometric to $G_2$.

**Proof.** As $G_1$ is isomorphic to $G_2$, there is a bijection $g : V_1 \to V_2$ such that
$\mu_{A_1}^-(x) = \mu_{A_2}^-(g(x)), \mu_{A_1}^+(x) = \mu_{A_2}^+(g(x))$ for all $x \in V_1$,
$\mu_{B_1}^-(xy) = \mu_{B_2}^-(g(x)g(y)), \mu_{B_1}^+(xy) = \mu_{B_2}^+(g(x)g(y))$ for all $x, y \in V_1$.

For each $u \in V_1$, we have

$$\delta_1^-(u,v) = \wedge \left\{ \sum_{i=1}^{n} \frac{1}{\mu_{B_1}^-(u_{i-1}u_i)} \right\} = \wedge \left\{ \sum_{i=1}^{n} \frac{1}{\mu_{B_2}^-(g(u_{i-1})g(u_i))} \right\} = \delta_2^-(g(u), g(v)),$$

$$\delta_1^+(u,v) = \wedge \left\{ \sum_{i=1}^{n} \frac{1}{\mu_{B_1}^+(u_{i-1}u_i)} \right\} = \wedge \left\{ \sum_{i=1}^{n} \frac{1}{\mu_{B_2}^+(g(u_{i-1})g(u_i))} \right\} = \delta_2^+(g(u), g(v)).$$

So, $G_2$ is isometric from $G_1$.

**Note 3.3.** (i) The above result is true even $G_1$ is co-weak isomorphic to $G_2$ also.
(ii) we know that, $G_1$ is isomorphic to $G_1$ implies $\overline{G_1}$ is isomorphic to $\overline{G_2}$. But this is not so in the case of isometry.

In the following example, we show that $G_1$ and $G_2$ are two interval-valued fuzzy graphs that $G_2$ is isometric from $G_1$ but $\overline{G_2}$ is not isometric from $\overline{G_1}$.

**Proposition 3.5.** Isometry on interval-valued fuzzy graphs is an equivalence relation.
**Proof.** Let $G_i = (A_i, B_i)$, $i = 1, 2, 3$ be the interval-valued fuzzy graphs with underlying sets $V_i$. Considering the identity map $i : V_1 \to V_1$, $G_1$ is isometric to $G_1$. Therefore isometry is reflexive.

To prove the symmetric, assume that $G_1$ is isometric to $G_2$. Hence $G_2$ is isometric from $G_1$ and $G_1$ is isometric from $G_2$. By rearranging, $G_2$ is isometric to $G_1$.

To prove the transitivity, let $G_1$ be isometric to $G_2$ and $G_2$ be isometric to $G_3$, i.e. $G_2$ is isometric from $G_1$ and $G_3$ is isometric from $G_2$. Then, for each $v \in V_1$, there exists a bijective map $g_v : V_1 \to V_2$ such that $\delta_1^-(v,u) = \delta_2^-(g_v(v), g_v(u))$,
$\delta_1^+(v,u) = \delta_2^+(g_v(v), g_v(u))$ for all $u \in V_1$.
Suppose that $g_v(v) = v'$. Similarly, for each $v' \in V_2$, there exists a bijective map

$h_{v'} : V_2 \to V_3$ such that $\delta_2^-(v',u') = \delta_3^-(h_{v'}(v'), h_{v'}(u'))$,

$\delta_2^+(v',u') = \delta_3^+(h_{v'}(v'), h_{v'}(u'))$ for all $u' \in V_2$. Now if $v \in V_1$,

$\delta_1^-(v,u) = \delta_2^-(g_v(v), g_v(u)) = \delta_2^-(v',u') = \delta_3^-(h_{v'}(v'), h_{v'}(u'))$

$= \delta_3^-(h_{v'}(g_v(v)), h_{v'}(g_v(u))) = \delta_3^-(h_{v'} \circ g_v(v), h_{v'} \circ g_v(u))$, for all $u \in V_1$.

$\delta_1^+(v,u) = \delta_2^+(g_v(v), g_v(u)) = \delta_3^+(h_{v'} \circ g_v(v), h_{v'} \circ g_v(u))$, for all $u \in V_1$.

Hence $G_3$ is isometric from $G_1$, using the composite map $h_{v'} \circ g_v : V_1 \to V_3$.

Thus isometry on interval-valued fuzzy graphs is an equivalence relation.

## 4. Conclusions

Graph theory is an extremely useful tool in solving the combinatorial problems in different areas including geometry, algebra, number theory, topology, operations research, optimization, computer science, etc. The interval-valued fuzzy sets constitute a generalization of the notion of fuzzy sets. The interval-valued fuzzy models give more precision, flexibility and compatibility to the system as compared to the classical and fuzzy models. In this paper, we define isometry on interval-valued fuzzy graphs and shown that isometry on interval-valued fuzzy graph is an equivalence relation.